\documentstyle[preprint,aps]{revtex}
\begin{document}
\tighten
\title{Non-radiative synthesis of $^7{\rm Be}$ in solar plasma}
\author{
       	N. V. Shevchenko$^{1,3}$, S. A. Rakityansky$^{1,2}$,
	S. A. Sofianos$^{2}$,\\  and \\V. B. Belyaev$^{1,4}$}
\address{
{\small 1)}
	Joint Institute  for Nuclear Research,Dubna, 141980, Russia\\
{\small 2)}
	Physics Department, University of South Africa,
	P.O.Box 392, Pretoria, South Africa\\
{\small 3)}
	Physics Department, Irkutsk State University, 
	Irkutsk 664003, Russia\\
{\small 4)}
	RCNP, Osaka University, Mihogaoko 10--1, Ibaraki,
	Osaka 567, Japan}
\date{\today}
\maketitle
\begin{abstract}
The nuclear reaction
	$e+{^3{\rm He}}+{^4{\rm He}}\rightarrow{}e +{^7{\rm Be}}$ 
is considered at thermonuclear energies. The motion of the 
electron is treated within the adiabatic approximation and the 
${^3{\rm He}}-{^4{\rm He}}$ scattering state is calculated using 
an effective ${^3{\rm He}}-{^4{\rm He}}$ potential constructed via 
the Marchenko inverse scattering method.
The reaction rate thus obtained for solar interior conditions is
approximately $10^{-4}$ of the corresponding rate for the radiative capture
${^3{\rm He}}+{^4{\rm He}}\rightarrow{}{^7{\rm Be}}+\gamma$. However, at the
conditions corresponding to the early stages of the primordial
nucleosynthesis, the non-radiative fusion is five times faster than the
radiative one. The importance of the non-radiative processes for the energy
balance in the sun, which for electrons is totally opaque, is discussed.
\\[0.3cm]
\end{abstract}
\section{Introduction}

In the standard model of the sun it is assumed that all nuclear
reactions forming the $pp$--chain, are generated by various two--body
nucleus--nucleus and electron--nucleus collisions with only one exception,
namely, the $e+p+p\rightarrow d+\nu$ reaction\cite{Bahcall}.
Other three--body reactions are ignored on the grounds that the triple
collisions occur less frequently than the binary collisions. 

Defining the collision as a situation in which the particles approach 
each other to a distance less than the range $R$ of their interaction, 
then the frequency of two--body collisions per unit volume is
$$
	       {\cal F}_{12}=\pi {R}^2v_{12}n_1n_2\ ,
$$
where $v_{12}$ is the relative velocity of particles 1 and 2, and
$n_i$ are their densities. The frequency of the triple collision 
${\cal F}_{123}$ is obtained if we multiply ${\cal F}_{12}$ 
by the probability of finding the particle 3 present when 1 and 2 are
scatterred  \cite{GW}, viz. by $(4\pi/3){R}^3n_3$. Then

$$
	       {\cal F}_{123}=\frac{4\pi^2}{3}{R}^5v_{12}n_1n_2n_3\ .
$$
From these elementary considerations one can find that triple collisions are
less probable than the binary ones. For example, in the centre of the
sun the densities of electrons, $^3$He, and $^4$He nuclei are (these
can be derived from the Tables 4.4 and 4.5 of Ref.\cite{Bahcall})
\begin{eqnarray}
\label{dens}
\nonumber
	n_e  &=&0.60\times 10^{26}\ {\rm cm}^{-3}\ ,\\
    	n_{{}^3{\rm He}} &=&0.23\times 10^{21}\ {\rm cm}^{-3}\ ,\\
\nonumber
   	n_{{}^4{\rm He}} &=&0.14\times 10^{26}\ {\rm cm}^{-3}\ ,
\end{eqnarray}
which together with a nuclear reaction range of ${R}\sim 10$\,fm and
a velocity of $v_{{}^3{\rm He}{}^4{\rm He}}\sim 3\times 10^7$\,cm/sec
corresponding to the temperature $1.56\times 10^7\ {}^{\circ}$\,K, 
give
\begin{eqnarray*}
      	{\cal F}_{{}^3{\rm He}{}^4{\rm He}}  &\sim & 0.3\times 10^{30}\
      	{\rm cm}^{-3}{\rm sec}^{-1}\ ,\\
      	{\cal F}_{e{}^3{\rm He}{}^4{\rm He}} &\sim & 0.8\times 10^{20}\
      	{\rm cm}^{-3}{\rm sec}^{-1}\ .
\end{eqnarray*}
This, however, is not a sufficient proof that the three--body reactions
can be ignored. This is due to the fact that not every collision 
results in a nuclear transition. The actual reaction rate
depends also on the transition probability which may be higher for 
three--body rather than for 
two--body initial states. This stems from the fact that triple 
collisions are kinematically less restricted than the binary ones. 
Some binary reactions are forbidden by conservation laws, e.g., 
conservation of angular momentum, parity, isospin etc. However,
such reactions could take place in the presence of a third particle.

Calculations of the three--body reaction rates performed so far
\cite{Picker,pde,pebe7}, corroborate this by showing that the 
simplified considerations  based on the frequency of collisions 
overestimate the differences between the two- and three-body reaction rates.
Thus, for example, in Ref.\cite{Picker} it was shown that under 
the conditions which prevailed during the primordial nucleosynthesis, 
the non-radiative capture $ e+n+p\to e+d$ occurs at a reaction rate which 
is five orders of magnitude less than the  corresponding
rate for the radiative capture $n+p\to \gamma +d$. In Refs.\cite{pde,pebe7}
the reaction rates for the non-radiative processes $e+p+d\to e+{}^3{\rm He}$
and $e+p+{}^7{\rm Be}\to e+{}^8{\rm B}$ were calculated. Again, it was found
that they are only four orders of magnitude less than for corresponding 
binary reactions  
		$p+d\to \gamma+{}^3{\rm He}$ 
and 
	      	$p+{}^7{\rm Be}\to \gamma+{}^8{\rm B}$. 
In contrast, a simplified estimation provides a rate  which  differs by 
ten orders of magnitude from them. 

The above examples clearly demonstrate that a reliable answer to the
question of whether or not a particular three--body reaction is negligible,
can be obtained only when  proper calculations are made. And surely in 
the case under consideration, such calculations should be done 
as we have  the  $^7$Be paradox on top of the long--standing solar 
neutrino problem.
This paradox originated from the combined analysis  of all experiments, in
which the neutrino flux from the sun was measured. This analysis came to the 
conclusion that the production of $^7$Be nuclei (more precisely, the flux of
neutrinos due to $^7$Be reactions) must be strongly suppressed or even be
negative \cite{Bah,Deg}. This implies that something is wrong either in the
standard model or in the experimental data. In this respect the fate of
$^7$Be in the  $pp$--chain is of  special interest.

According to the standard model this nucleus is produced via the radiative
fusion
\begin{equation}
\label{347g}
		{}^3{\rm He}+{}^4{\rm He}\
	        \longrightarrow\ {}^7{\rm Be}+\gamma\ ,
\end{equation}
in which the released energy (1.59\,MeV) is carried away by the photon.
Another possibility for a $^7$Be production is  the three--body 
non-radiative fusion
\begin{equation}
\label{e347e}
	     	e+{}^3{\rm He}+{}^4{\rm He}\
	     	\longrightarrow\ e + {}^7{\rm Be}\ .
\end{equation}
This is a kind of Auger transition in which the same amount of the released
energy is carried away by the electron.

As we have seen, the triple collisions (\ref{e347e}) are less frequent than
the binary ones (\ref{347g}) and thus one can argue that they are
unimportant. We have, however, at least three reasons to perform
explicit calculations for the reaction rate of the process (\ref{e347e}).
Firstly, since the neutrino paradox persists we have to be scrupulous 
in examining any conjecture on the matter. Secondly, although the 
three-body reaction may be unimportant in the neutrino problem,
it may contribute to the energy balance in the sun since
 the energy released in each non-radiative fusion event remains in the
center of the sun because the electron which carries it, cannot escape.
In contrast to  photons, for electrons the sun is totally opaque. To
estimate this additional heating, we need to know the three--body reaction
rate. Finally, at the early stages of the primordial nucleosynthesis 
the temperature as well as the particle densities were much higher 
than they are in the solar plasma. We therefore may expect that the 
three--body reactions were important at that times.

The paper is organized as follows. In Sec. II we describe our formalism in
general and in Sec. III we outline the procedure and approximations 
employed to evaluate the transition operator.  Sec. IV is devoted to 
the potentials we use to describe the interactions among the particles, 
and In  Sec. V we present our results and conclusions.
%
\section{Three--body reaction rate}
The nuclei $^3$He  and $^4$He are stable clusters (in what follows we denote
them by $\tau$ and $\alpha$ respectively) in the low--lying states
of $^7$Be (see Ref. \cite{be7spec} and references therein). We can,
therefore, reduce the eight--body problem $e(NNN)(NNNN)$ to the three--body
one, namely, $e-\tau-\alpha$.  The corresponding Jacobi vectors in
configuration and momentum spaces are shown in Fig. \ref{Jacobi}.

\begin{figure}
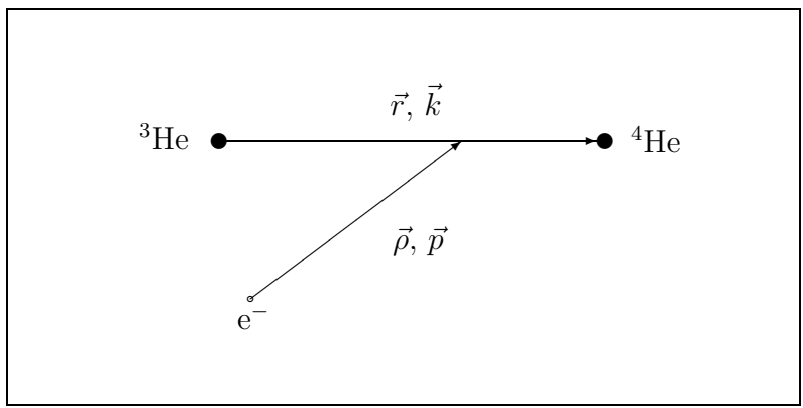

\caption{Low--lying energy levels  (in MeV) of $^7$Be nucleus and the
$^3$He-$^4$He threshold energy.}
\label{spectrum}
\begin{center}
\unitlength=0.6mm
%
\end{center}
\end{figure}

Let ${\cal R}({\bf k},{\bf p}\to{\bf p'})$ be the reaction rate
per unit volume per second for the collision process (\ref{e347e}) which
starts with the momenta $({\bf k},{\bf p})$ and ends up with
${\bf p'}$.  The plasma particles are in a thermodynamical equilibrium and
their momenta are distributed according to Maxwell's law
\begin{eqnarray}
\nonumber
    N_{\bf k}(\theta)&=&(2\pi\mu\kappa\theta)^{-3/2}
    \exp\left(-\displaystyle
    \frac{k^2}{2\mu\kappa \theta}\right)\,,\\
\nonumber
    N_{\bf p}(\theta)&=&(2\pi m\kappa \theta)^{-3/2}
   \exp\left(-{\displaystyle \frac{p^2}{2m\kappa \theta}}\right)\,,
\end{eqnarray}
where $N_{\bf k}$ and $N_{\bf p}$ are the probability densities, $\mu$
is the $\tau\alpha$--reduced mass, $m$ is the electron mass,
$\kappa$ is the Boltzmann constant, and $\theta$ is the plasma temperature.
We are cocerned with the total rate of the transition from an initial 
state with any $({\bf k},{\bf p})$ to a final state with all possible 
${\bf p'}$. Thus the reaction rate ${\cal R}({\bf k},{\bf p}\to{\bf p'})$
must be  averaged over the initial momenta ${\bf k}$ and ${\bf p}$ and
integrated over the final momentum ${\bf p'}$, i.e.,
\begin{equation}
\label{RA}
       	\langle{\cal R}\rangle_\theta=\int\int\int d{\bf k} d{\bf p}
       	d{\bf p}' \,{\cal R}({\bf k},{\bf p}
       	\to{\bf p}')N_{\bf k}(\theta)N_{\bf p}(\theta)\,.
\end{equation}
Similarly to  the two--body reaction theory where the average reaction 
rate $\langle{\cal R}_{12}\rangle_\theta$ is written as a product 
of  $\langle \sigma_{12}v_{12}\rangle_\theta$
(which is referred to as the reaction rate per particle pair) and the
particle densities  \cite{RR},
$$
        \langle{\cal R}_{12}\rangle_\theta=
        n_1n_2\langle \sigma_{12}v_{12}\rangle_\theta \ ,
$$
the three--body reaction rate $\langle{\cal R}\rangle_\theta$ can also be
factorized in the same manner,
\begin{equation}
\label{factor}
         \langle{\cal R}\rangle_\theta= n_e n_{\tau} n_{\alpha}
         \langle{\Sigma}\rangle_\theta\ .
\end{equation}
Indeed, using the  general formula for a three--body reaction rate
\cite{GW} we obtain
\begin{eqnarray}
\label{odin}
\nonumber
     	{\cal R}\,({\bf k},{\bf p} \to {\bf p'})&=&\frac{(2 \pi)^7}{2}
     	n_{e} n_{\tau} n_{\alpha}\sum\limits_{n m_s m_j}
     	\delta\left(\frac{p'{\,}^2}{2m}- E^{(n)} -
     	\frac{k^2}{2\mu} - \frac{p^2}{2m} \right)\\
&&\\
\nonumber
        &\times& |\, \langle
	\psi_{7\,jm_j}^{(n)};{\bf p'}\,|\, T\,|\, 
	{{\bf k},m_s};{\bf p}\rangle  |^2\ ,
\end{eqnarray}
where the  $\delta$--function secures the energy conservation for 
the transition to the $n$-th level of  $^7$Be nucleus (see Fig. 
\ref{spectrum}) accompanied by the energy release $E^{(n)}$
and $T$ is the transition operator for the process (\ref{e347e}). 
The initial and final asymptotic state wave functions for this reaction, 
		$|{\bf k},m_s;{\bf p}\rangle$ 
and
		$|\psi_{7\,jm_j}^{(n)};{\bf p'}\rangle$,
are characterized, apart from the momenta, by the nuclear spin 
$s=1/2$  of the $^3$He, the total angular momentum  $j$
of $^7$Be (which depends on the state $n$),
and their third  components $m_s$ and $m_j$  respectively. 
They are normalized as
\begin{eqnarray*}
      	\langle \, {\bf k}',m_s'; {\bf p'} \,\,|\,\,
	{\bf k},m_s; {\bf p} \rangle &=& \delta_{m_{s}'m_s}
      	\delta ({\bf k}-{\bf k'}) \,\,\delta ({\bf p}-{\bf p'})\ ,\\
      	&&\\
	\langle \,\psi_{7\,jm_j'}^{(n')};{\bf p'} \,\,|\,\,
    	\psi_{7\,jm_j}^{(n)};{\bf p}\rangle &=& \delta_{n'n}\,
	\delta_{m_j'm_j} \,\delta ({\bf p}-{\bf p'})\ .
\end{eqnarray*}
The sum $\frac12\sum_{m_s m_j}$ in (\ref{odin}) is for
averaging over the initial and summing over the  final spin orientations.
Therefore the quantity $\langle{\Sigma}\rangle_\theta$ defined by
Eq. (\ref{factor}), in analogy to the two--body case,  may be  referred 
to as the reaction rate per particle trio and is given by
\begin{eqnarray}
\label{trio}
\nonumber
         \langle{\Sigma}\rangle_\theta &=&
       	\displaystyle
       	\frac{(2\pi)^7}{2}\sum\limits_{n m_s m_j}
       	\int\int\int d{\bf k} d{\bf p} d{\bf p}'\,
     	\delta\left(\frac{p'{\,}^2}{2m}- E^{(n)} -
     	\frac{k^2}{2\mu} - \frac{p^2}{2m} \right) \\
     	&&\\
\nonumber     	
	&&\times\displaystyle {\bf |}\, \langle
	\psi_{7\, jm_j}^{(n)};{\bf p'}\,|\, T\,|\, {\bf k},
	m_s;{\bf p}\rangle \,{\bf |}^2\,
       N_{\bf k}(\theta)N_{\bf p}(\theta)\,.
\end{eqnarray}
In the next two sections we describe how the various ingredients needed 
to obtain the matrix element 
	$\langle \psi_{7\, j m_j}^{(n)};{\bf p'}\,|\, T\,|\,
	{{\bf k},m_s};{\bf p} \rangle$ 
were calculated.
%
\section{Transition operator}
The three--body quantum state $|e+ {}^3{\rm He}+ {}^4{\rm He} \rangle$ 
belongs to the Hamiltonian
\begin{equation}
\label{Ham}
             H=H_0+h_0+V_N+V_e\,,
\end{equation}
where $H_0$ and $h_0$ are the kinetic energy operators associated with the
Jacobi variables ${\bf r}$ and {\boldmath $\rho$} respectively
(see Fig. \ref{Jacobi}),
$$
     V_N=V_{\tau\alpha}^{(s)}+V_{\tau\alpha}^{(c)}
$$
is the  nuclear $\tau$-$\alpha$
potential, which includes strong and Coulombic parts. The $V_e$ consists
of the $e$-$\tau$ and $e$-$\alpha$  Coulomb potentials
$$
     V_e=V_{e\tau}+V_{e\alpha}\ .
$$
The various transitions in this three--body system are determined by
the relevant matrix elements of the $T$--operator obeying the
Lippmann--Schwinger equation
\begin{equation}
\label{LSeq}
            T(z)=V_N+V_e +(V_N+V_e)\frac{1}{z-H_0-h_0}T(z)\,.
\end{equation}
The probability of the transition (\ref{e347e}), i.e. of
\begin{equation}
\label{asstran}
	\left|{{\bf k},m_s};{\bf
	p}\right\rangle \ \stackrel{V_N+V_e}{\Longrightarrow} \
	\left|\psi_{7\,jm_j}^{(n)};{\bf p'}\right\rangle \ ,
\end{equation}
is therefore determined from the matrix element
	$\langle \psi_{7\,jm_j}^{(n)};{\bf p'}\,|\, T\,|\,
	{{\bf k},m_s};{\bf p}\rangle $.

Making use of the special conditions prevailing in solar plasma and 
the smallness of the electron mass, we can significantly simplify 
Eq. (\ref{LSeq}). Indeed, the average kinetic energy of the particles 
in the plasma, $<E^{kin}>\sim\kappa\theta\approx 1$\,keV,  is the same 
for nuclei and electrons, but the velocity of an electron is three 
orders of magnitude higher than that of $^3$He or $^4$He. Therefore, 
while the $\tau$ approaches $\alpha$  very slowly, the electron dashes 
nearby picking up the energy and leaving the heavy particles in a 
bound state. Due to this we can treat the relative $\tau\alpha$ motion 
adiabatically. For  a three--body collision such an approximation 
results in two  simplifications.  In the first one 
the action of the potentials $V_N$ and $V_e$ can be separated. 
Indeed, while the electron starts from its asymptotic state 
$|{\bf p}\rangle$, the heavy particles have already interacted 
via the potential $V_N$ and formed the two--body scattering state
\begin{equation}
\label{scn}
	\left|{\bf k},m_s\right\rangle
	\ \stackrel{V_N}{\Longrightarrow} \
	\left|\psi_{{\bf k},m_s}\right\rangle\ .
\end{equation}
Therefore, instead of the transition (\ref{asstran}) caused by both $V_N$
and $V_e$, in the adiabatic approximation we may consider the transition
\begin{equation}
\label{distort}
	\left|\psi_{{\bf k},m_s};{\bf p}\right\rangle
	\ \stackrel{V_e}{\Longrightarrow} \
	\left|\psi^{(n)}_{7\,jm_j};{\bf p'}\right\rangle\ ,
\end{equation}
where the interaction $V_N$ is taken into account by (\ref{scn}).
In other words, the transition (\ref{asstran})  effectively  occurs in  
two steps, (\ref{scn}) and (\ref{distort}). 
As a result the two--body scattering problem (\ref{scn}) can be solved
separately. 

In the second  simplification the transition (\ref{distort})   may 
be described using the fixed scatterer $T$--matrix defined as
\begin{equation}
\label{LSeq1}
	\tilde T(z)=V_e+V_e\frac{1}{z-h_0}\tilde T(z)\ .
\end{equation}
Within the above approximation we have
\begin{equation}
\label{Tappr}
	\langle\psi^{(n)}_{7\,jm_j};{\bf p'}|T|
	{\bf k},m_s;{\bf p}\rangle\approx
	\langle\psi^{(n)}_{7\,jm_j};{\bf p'}|
	\tilde T|\psi_{{\bf k},m_s};{\bf p}\rangle\ .
\end{equation}
A further simplification can be achieved when  Eq. (\ref{LSeq1}) is
solved iteratively. For the solar plasma electrons one
has the condition $e^2/hv < 1$, which is a sufficient to treat Coulomb
interactions in Eq.~(\ref{LSeq1}) perturbatively, i.e.,
$$
	\tilde T(z)=V_e+V_e\frac{1}{z-h_0}V_e+
	V_e\frac{1}{z-h_0}V_e\frac{1}{z-h_0}V_e
	+\cdots\, ,
$$
Furthermore, the average potential energy of the electron is of atomic
order of magnitude, $<V_e>\sim 10\ {\rm eV}$, while its kinetic energy
in solar plasma is two orders of magnitude higher,
\mbox{$<E^{kin}_e>\sim 1\ {\rm keV}$}, which implies that the above
iterations should converge very fast. Therefore, we may retain
only the first (Born) term \cite{Taylor}.

Finally, the exact matrix elements of the $T$--operator
can be replaced by the  approximate ones
\begin{equation}
\label{dwba}
        \langle \psi_{7\,jm_j}^{(n)};{\bf p'}\,|\, T\,|\,
	{{\bf k},m_s};{\bf p}\rangle \approx
        \langle \psi_{7\,jm_j}^{(n)};{\bf p'}\,|\, V_e\,|\,
	\psi_{{\bf k},m_s};{\bf p}\rangle\ .
\end{equation}
This can be considered as a three-body generalization of
the distorted wave Born approximation (DWBA) which is widely used
in the theory of the two--body nuclear reactions.
%
\section{Potentials}
The evaluation of the matrix element (\ref{dwba}) requires the knowledge of
the potential $V_e$ and the wave functions $\psi_{{\bf k},m_s}({\bf r})$
and $\psi_{7\,jm_j}^{(n)}({\bf r})$. In order to obtain the wave functions we
solved the $\tau\alpha$ two--body problem using the Jost function method,
proposed in Refs. \cite{RSA96,SR97}, with the $\tau\alpha$ effective
potential $V_{\tau\alpha}=V_{\tau\alpha}^{(s)}+V_{\tau\alpha}^{(c)}$.
%
\subsection{Coulomb forces}
The electron as  a point-like particle generates a pure Coulomb field. 
However, the electric charges of $\tau$ and $\alpha$ are 
distributed within their nuclear volumes which have a size comparable to the 
typical range of the nuclear reaction (few fm). Furthermore, the
average collision energy, $\sim 1$\,keV,  is comparable to the 
height of the Coulomb barriers and thus the charge distributions 
should be taken into account. This is achieved by assuming that the 
nuclear charge distributions are  spherically symmetric and have
a Gaussian form
\begin{equation}
\label{qd}
  	q_i(\varrho)=2e\left(\frac{3}{2\pi\langle R_i^2\rangle}
	\right)^\frac32
  	\exp\left(-\frac{3\varrho^2}{2\langle R_i^2\rangle}\right)\ ,
  	\qquad i=\tau,\alpha\ ,
\end{equation}
where $\varrho$ is the distance from the centre of the nucleus. The
normalization coefficient and the slope parameter in (\ref{qd}) are 
chosen to give the total charges and the r.m.s. radii of the nuclei, 
i.~e.
$$
     	\int q_i(\varrho)\,d^3\!\varrho=2e\ ,
$$
$$
	\frac{1}{2e}\int \varrho^2 q_i(\varrho)\,d^3
	\!\varrho=\langle R_i^2\rangle\ .
$$
For the r.m.s. radii we used the experimental values \cite{rtau,ralfa}
$$
	  \sqrt{\langle R_{\tau}^2\rangle}=1.93\, {\rm fm}\ ,\qquad
          \sqrt{\langle R_{\alpha}^2\rangle}=1.67\, {\rm fm}\ .
$$
Using the space charge distribution (\ref{qd}), we obtain
\begin{equation}
\label{Vei}
          V_{ei}(\varrho)=\frac{2e^2}{\varrho}\,{\rm erf}\left(
          \varrho\sqrt{\frac{3}{2\langle R_i^2\rangle}}\,\right)\ ,
          \qquad i=\tau,\alpha\ ,
\end{equation}
\begin{equation}
\label{tac}
 	V_{\tau\alpha}^{(c)}(\varrho)= \frac{4 e^2}{\varrho} \,
	\sqrt{\frac{6}{\pi\langle R_{\alpha}^2\rangle}}
	\int_0^\infty\,\exp\left[-\frac{3({\varrho'}^2
	+\varrho^2)}{2\langle R_{\alpha}^2\rangle}
	\right]\sinh\left(\frac{3}
	{\langle R_{\alpha}^2\rangle}\varrho\varrho'\right) \,
	{\rm erf}\left(\varrho'\sqrt{\frac{3}{2\langle 
	R_{\tau}^2\rangle}}\,\right)\,d\varrho'\ .
\end{equation}
At large distances this electron--nucleus potential (\ref{Vei})
has the usual  Coulomb tail $2e^2/\varrho$. At short
distances, however, it is essentially different without having 
a singularity at $\varrho=0$. The same can be said about the 
nucleus--nucleus interaction (\ref{tac}).

The electron which participates in the three--body reaction, is only one of
the great number of the plasma electrons. All the others are spectators
surrounding the nuclei and thus their combined Coulomb field
reduces  the electric fields of the nuclei. In the standard
Debye--H\"uckel theory \cite{salpeter} such screening effect is taken into
account by introducing an exponentially decaying factor into the Coulomb
part of the nucleus--nucleus potential.
Following this approach, we replace the potentials (\ref{Vei}) and
(\ref{tac}) by the screened ones
\begin{eqnarray}
\label{sVei}
	   V_{ei}(\varrho)\ &\stackrel{screen}{\longrightarrow}&
	   V_{ei}(\varrho)\exp\left(-\frac{\varrho}{D}\right)\ ,
		   \qquad i=\tau,\alpha\ ,\\
\nonumber
&&\\
\label{stac}
	V_{\tau\alpha}^{(c)}(\varrho)\ &\stackrel{screen}{\longrightarrow}&\
	V_{\tau\alpha}^{(c)}(\varrho)\exp\left(-\frac{\varrho}{D}\right)\ ,
\end{eqnarray}
with the Debye radius being $D=21800$\,fm which corresponds to the 
solar plasma conditions and is typical for other stars as well \cite{leeb}.

%
\subsection{Nuclear forces}
%
As  can be seen in Fig. \ref{spectrum}, when $^3$He and $^4$He nuclei
are fused in solar plasma, they may form either the ground or the first 
excited state of $^7$Be, the quantum numbers $j^\pi$ being 3/2$^-$ and 1/2$^-$ 
respectively. The rest of the excited states are situated very high 
and hence we can safely ignore the virtual transitions via them. 
Therefore, in constructing the nucleus--nucleus scattering wave function
$\psi_{{\bf k},m_s}$ we assume that $^3$He and $^4$He
may interact if their total angular momentum $j\le 3/2$. This restricts
the allowable values of their relative orbital momentum $\ell$ 
to $\ell\le 2$. The state with $\ell=2$  can also be ignored 
since at  collision energies $\sim 1$\,keV the  contribution from
the higher partial waves  diminishes very fast. Thus for
the nuclear forces between $^3$He and $^4$He we consider those 
corresponding to the lowest 
three ($\ell, j$)--states, namely, (0,1/2), (1,1/2), and (1,3/2).

To construct the  corresponding $\tau\alpha$--potentials, we employ the
Marchenko inverse scattering method \cite{March,CS77} which is briefly
outlined below. Within this method we
obtained energy independent local potentials $V_{\tau\alpha}^{(s)}(r)$ 
for each of the above three partial waves, which reproduce the 
the Resonating Group Model (RGM) $\tau\alpha$--scattering phase--shifts 
\cite{RGM1,RGM2} and give the correct binding energies for  the ground 
and first excited states of $^7$Be (in the $\tau\alpha$--model).

In the Marchenko inverse scattering method a unique, energy--independent,
$\ell$--dependent, local potential $V_\ell(r)$ is obtained from
\begin{equation}
      V_\ell(r) = - 2\frac{{\rm d}}{{\rm d}r} K_\ell(r,r)
\label{velp}
\end{equation}
where the kernel $K_\ell(r,r')$ obeys the Marchenko  fundamental
equation
\begin{equation}
     K_\ell(r,r') + F_\ell(r,r') +\int_r^\infty
     K_\ell(r,r'') F_\ell(r'',r')dr'' = 0.
\label{march}
\end{equation}
The driving term $F_\ell(r,r')$ is given by
\begin{equation}
       F_\ell(r,r') = \frac{1}{2\pi}
          \int_{-\infty}^{+\infty} w_\ell^+(kr)
          [1-S_\ell(k)] w_\ell^+(kr'){\rm d}k
          + A_\ell w_\ell^+(b_{\ell}r) w_\ell^+(b_{\ell}r').
\label{fl}
\end{equation}
$S_\ell(k)$ is the S--matrix for the specific partial wave,
and the function $w_\ell^+(z)$ is related to the spherical
Hankel function $h_\ell^{(+)}(z)$ by
\begin{equation}
       w_\ell^+(z) = i e^{i\pi\ell}z h_\ell^{(+)}(z). \label{wl}
\end{equation}
Furthermore, $A_\ell$ is the so--called asymptotic bound state
normalisation constant, while
$b_{\ell} = \sqrt{ - 2\mu E_{b}^{\ell}}$, $E_{b}^{\ell}$
being the $\tau\alpha$ bound state energy and $\mu$ their reduced mass.

The evaluation of $F_\ell(r,r')$ requires the knowledge of the 
S--matrix for all real energies from the elastic scattering 
threshold to infinity, together with the reflection property 
$S_\ell(- k)= 1/S_\ell(k)$,  as well as the
binding energies and the corresponding asymptotic bound state 
normalisation constants. It is greatly simplified by choosing 
a rational (Bargmann--type) parametrisation
\begin{equation}
     S_\ell(k) = \prod_{m=1}^{N_B}\frac{k+ib_\ell^m}{k-ib_\ell^m}
         \prod_{n=1}^{N_{\ell}}
     \frac{k+\alpha_{n}^{\ell}}{k-\alpha_{n}^{\ell}} \label{sl}
\end{equation}
where $N_B$ is the number of  bound states present.
The number $N_\ell+N_B$ must be even to satisfy the properties 
of the $S$--matrix.
The $\alpha_{n}^{\ell}$ are complex numbers used to fit
the (numerically) given S--matrix. We mention here that the 
so--called Pauli Forbidden States (PFS) are
ignored. This implies that the resulting  potentials are
the unique supersymmetric shallow \cite{Alt1} partners sustaining 
only the physical  bound states.
  
With the choice (\ref{sl}) the integration in Eq. (\ref{fl})
can be easily performed analytically (for more details see
\cite{CS77,Alt1,Alt2}).

\begin{figure}
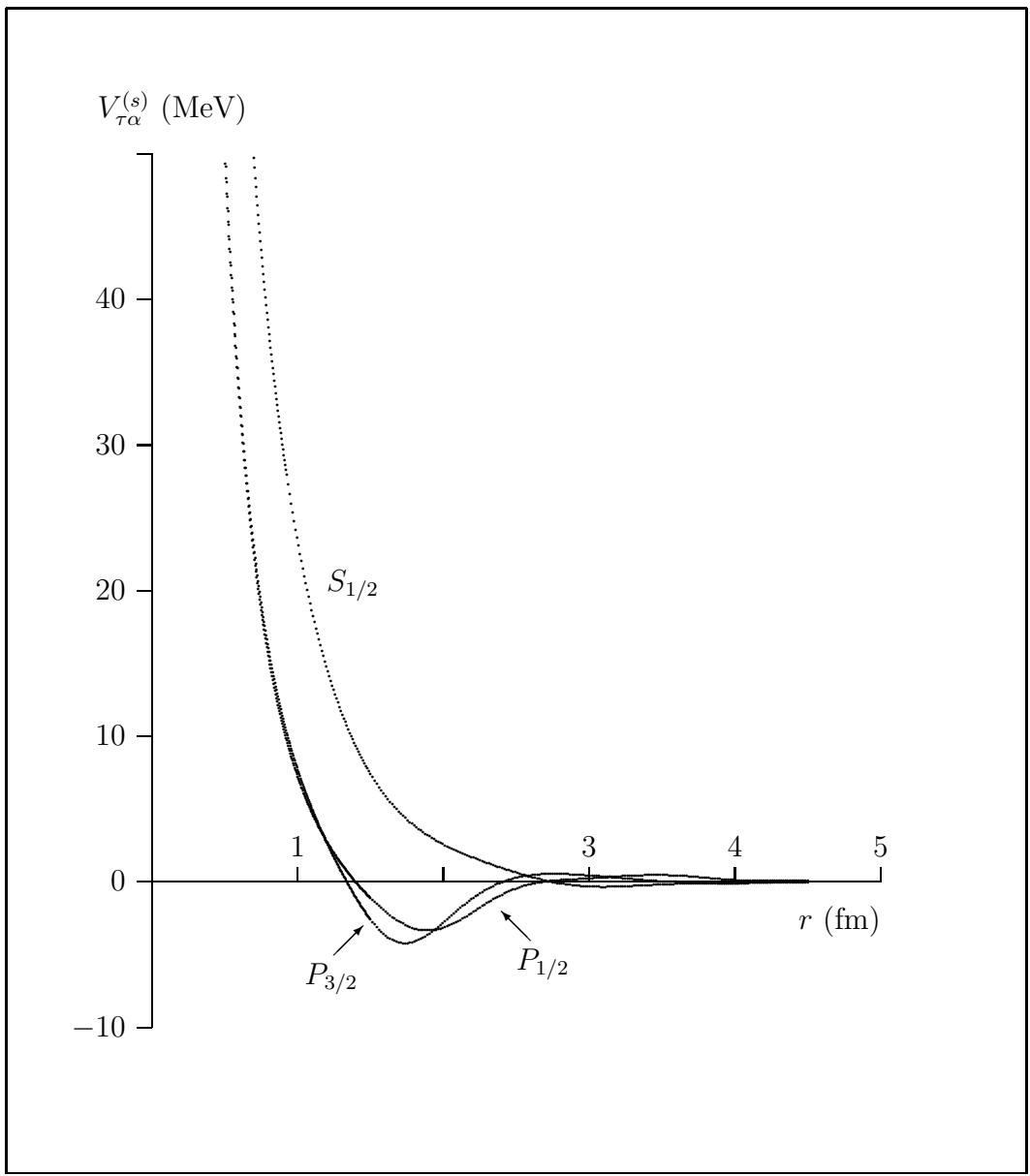

\caption{Effective $^3$He-$^4$He potential for three different states
corresponding to  relative orbital angular momentum $\ell=0,1$ and 
total angular momentum $J=1/2,\,3/2$.}
\label{pots}
\begin{center}
\unitlength=0.2mm
%
}
\end{picture}
\end{center}
\end{figure}

The potentials $V_{\tau\alpha}^{(s)}$ thus obtained for the 
$\tau\alpha$  system  in the channels S$_{1/2}$, P$_{1/2}$, 
and P$_{3/2}$ are shown in Fig. \ref{pots}.
%
\section{Results and conclusions}
The plasma temperature in the center of the sun is
$\theta_0=15.6\times 10^6\, {}^{\circ}$K \cite{Bahcall} and decreases,
by a factor of ten  towards the surface. Calculating the reaction rate for
the process (\ref{e347e}) we need, therefore, to cover this range of $\theta$.
The same $pp$--chain reactions occur in other stars as well with 
similar conditions. In the blue stars, which are hotter than the sun, 
the plasma temperature is much
higher. Thus, it is worthwhile to extend our calculations beyond the solar
temperatures too. Another reason for such an extension comes from the
necessity to examine the role of the triple collisions in the primordial
nucleosynthesis which also proceeded via the $pp$--cycle. The temperature
at which the creation of the nuclei occured in the early universe was
$\sim 10^9\, {}^{\circ}$K \cite{RR}.

We, therefore, calculated the three--body reaction rate for the values of
$\theta$ ranging from  $1\times 10^6\,{}^{\circ}$K to $3\times 
10^9\,{}^{\circ}$K. The results of our calculations are given in Table
\ref{trate}. Since the reaction rate obtained can be used  not only 
in  models of the sun but also for other stellar objects as
well where the plasma densities may be quite different, we present it in
units of ${\rm cm}^6{\rm mole}^{-2}{\rm sec}^{-1}$ which are convenient for
a general use \cite{fowler}. The meaning of these units is that instead of
$n_e n_{\tau} n_{\alpha}$  we multiply $\langle\Sigma\rangle$ by $N_A^2$,
i.e. the Avogadro number squared. Then for any specific densities the
reaction rate, in the units ${\rm cm}^{-3}{\rm sec}^{-1}$, can be obtained 
from Table \ref{trate} by multiplying the values given there, by
$n_e n_{\tau} n_{\alpha}/N_A^2$.

To answer the question of the role played by the three--body reaction
(\ref{e347e}) in the solar $pp$--cycle, we need to compare its reaction 
rate with the corresponding rate of the two--body  process (\ref{347g}). 
Using the particle densities (\ref{dens}), we obtain for the center of the
sun
$$
     	\langle{\cal R}_{e\tau\alpha}\rangle_{\theta_0}=
     	7.15\times10^{3}\ {\rm cm}^{-3}{\rm sec}^{-1}\ ,
$$
while for the corresponding two--body process \cite{fowler}
$$
    	\langle{\cal R}_{\tau\alpha}\rangle_{\theta_0}=
    	2.44\times10^{7}\ {\rm cm}^{-3}{\rm sec}^{-1}\ .
$$
Their ratio is rather small,
\begin{equation}
\label{rat0}
    \frac{\langle{\cal R}_{e\tau\alpha}\rangle_{\theta_0}}
    {\langle{\cal R}_{\tau\alpha}\rangle_{\theta_0}}=
    2.94\times10^{-4}\ ,
\end{equation}
but not as small as one would guess from the simplified considerations
discussed in the introduction. The reaction rate
$\langle{\cal R}_{e\tau\alpha}\rangle_{\theta}$ in other parts of the sun
together with its ratio to the two--body reaction rate are
given in Table \ref{rrate}. In obtaining these data, we used the
dependences of $n_e$, $n_{\tau}$, $n_{\alpha}$, and $\theta$ on the distace
$R$ from the solar centre given in Ref. \cite{Bahcall}.

When we move away from the center of the sun the temperature and the 
electron density decrease. As a result
the ratio $\langle{\cal R}_{e\tau\alpha}\rangle_{\theta}/
\langle{\cal R}_{\tau\alpha}\rangle_{\theta}$ becomes even smaller than
(\ref{rat0}). But the nuclear burning occurs mainly in the central zone
of the sun ($R\sim0.1 R_\odot$). For example, the two--body reaction
rate $\langle{\cal R}_{\tau\alpha}\rangle_{\theta}$ at the radius
$0.1 R_\odot$ is already one order of magnitude less than at the centre
\cite{fowler}.
 Meanwhile, within the distance $R\lesssim0.1 R_\odot$ from 
the solar centre
the ratio $\langle{\cal R}_{e\tau\alpha}\rangle_{\theta}/
\langle{\cal R}_{\tau\alpha}\rangle_{\theta}$ remains practically the same,
$\sim10^{-4}$. Therefore, in calculating the
network of the solar $pp$--chain reactions one can ignore the non-radiative
synthesis of $^7$Be only if the required accuracy is less than
$\sim 0.01\,\%$.

For more hot and dense stars such accuracy restrictions 
are more essential as the ratio
\begin{equation}
\label{ratio}
    \frac{\langle{\cal R}_{e\tau\alpha}\rangle_{\theta}}
    {\langle{\cal R}_{\tau\alpha}\rangle_{\theta}}=
    \frac{\langle\Sigma_{e\tau\alpha}\rangle_{\theta}}
    {\langle\sigma_{\tau\alpha}v_{\tau\alpha}\rangle_{\theta}}n_e
\end{equation}
is proportional to the electron density $n_e$. Furthermore, by
comparing the data of Table \ref{trate} with the corresponding data for 
the two--body reaction given in Ref. \cite{fowler}, one can find that
the ratio $\langle\Sigma_{e\tau\alpha}\rangle_{\theta}/
\langle\sigma_{\tau\alpha}v_{\tau\alpha}\rangle_{\theta}$ increases with
temperature.

Another theory in which the omission of the three--body non-radiative 
synthesis of $^7$Be may be detrimental is that of the primordial 
nucleosynthesis during which practically all the nuclei were created 
when the temperature was in the range
\begin{equation}
\label{range}
		0.3\times10^9\ {}^{\circ}{\rm K}
		<\theta<3\times10^9\ {}^{\circ}{\rm K}
\end{equation}
with the electron density being (see Ref. \cite{Picker})
\begin{equation}
\label{nepair}
		n_e\approx 1.5\times10^{29}\left(\frac
		{\theta}{10^9}\right)^{\frac32}\exp\left(
		-\frac{5.93\times10^9}{\theta}\right)\ {\rm cm}^{-3}\ .
\end{equation}
At $\theta=10^9\ {}^{\circ}{\rm K}$, for example, formula (\ref{nepair})
gives
$$
	n_e\approx 0.40\times10^{27}\ {\rm cm}^{-3}\ , \qquad
	\theta=10^9\, {}^{\circ}{\rm K}\ ,
$$
which is one order of magnitude higher than the electron density
(\ref{dens}) in the solar interior.

Using the data of Table \ref{trate} and the corresponding data  for the
two--body reaction (\ref{347g}) of Ref. \cite{fowler}, together with the
electron density (\ref{nepair}), we calculated the ratio (\ref{ratio}) for
the primordial nucleosynthesis. The temperature dependence of this ratio
which we found for the temperature range (\ref{range}) is given in
Table \ref{prim}.  It is seen that at the earliest stages of the
nucleosynthesis, when the temperature was $\theta\approx 3\times10^9\
{}^{\circ}{\rm K}$, the non-radiative synthesis (\ref{e347e}) of $^7$Be
nuclei was five times faster than for the two--body fusion (\ref{347g}).
Thus, if the electron density given by formula (\ref{nepair}) is not far 
from the actual density that existed at the early stages of the Universe,
then three--body reactions of the type (\ref{e347e}) must be taken 
into account in  theories concerning primordial nucleosynthesis.
It can make a differences since the uncertainty in the $^7$Be 
production  in the primordial nucleosynthesis is currently assumed 
to be of the order of 16\% \cite{sarkar}.

Despite the fact that in the solar plasma the contribution of the 
three--body process is rather small, in enegy balance considerations 
this might not be the case. In each event (\ref{e347e}) or (\ref{347g}) 
an energy of 1.59\,MeV is released. The energy produced by the radiative 
capture (\ref{347g}), is carried away by a photon which can penetrate 
the inner layers of the sun and even escape. In contrast, the electron 
which receives the energy produced in the reaction  (\ref{e347e})
remains in the interior of the sun and  eventually 
distributes its excess  energy among the other plasma particles. This
causes an additional heating which is not taken into account by the standard
model of the sun. 

To evaluate the importance of three--body reactions in the energy balance 
quantitatively we calculated the rates of the energy released, due to 
the reactions (\ref{e347e})
per cm$^3$ per second at different radial positions in the sun. These are 
presented in Table \ref{rrate}. These results were obtained using the 
reaction rates of Table \ref{trate} together with the radial 
dependences of the particle densities and the temperature given in 
Tables 4.4 and 4.5 of Ref.\cite{Bahcall}. By integrating 
over the sun volume, we obtain the total energy generated in the sun 
every second from  the non-radiative synthesis of $^7$Be,
$$
       \left(\frac{{\rm d}{\cal E}}{{\rm d}t}\right)_{e\tau\alpha}=
       0.582\times 10^{16}\,\ \frac{{\rm erg}}{{\sec}}\ .
$$
As compared to the total photon luminosity of the sun \cite{Bahcall}
$$
       L_\odot=0.386\times10^{34}\,\ \frac{{\rm erg}}{{\sec}}
$$
it is rather small. Even during  the whole period of its existence,
$T_\odot\approx 4.55\times 10^9$ years, the sun has generated, via the
three--body reaction (\ref{e347e}),
$$
       \left(\frac{{\rm d}{\cal E}}{{\rm d}t}\right)_{e\tau\alpha}\,
       T_\odot \approx 0.835\times 10^{33}\,\ {\rm erg}\ ,
$$
which is less than the energy radiated every second. This energy,
however, has been  stored inside the sun unless it
was somehow transferred to the surface. In this connection the energy balance
in the sun should, perhaps, be re-examined with inclusion not only the
reaction (\ref{e347e}) but also other nonradiative fusion reactions such as
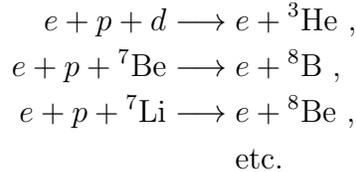
\begin{eqnarray*}
		e+p+d   &\longrightarrow&e+ {}^3{\rm He}\ ,\\
     e+p+{}^7{\rm Be}   &\longrightarrow&e+ {}^8{\rm B}\ ,\\
     e+p+{}^7{\rm Li}   &\longrightarrow&e+ {}^8{\rm Be}\ ,\\
			&& {\rm etc.}
\end{eqnarray*}
In conclusion, we can say that the three--body reaction (\ref{e347e})
contributes to $^7$Be production in the sun only  about 0.01\%, it perhaps
plays a role in the energy balance in the sun, and definitely it was
important at the early stages of the primordial nucleosynthesis.

\bigskip
\noindent
{\Large \bf Acknowledgements}\\
Financial support from the Russian Foundation for Basic Research
(grant \# RFBR 96 - 02 - 18678) and
the Foundation for Research Development of
South Africa is greatly appreciated. One of us (VBB) expresses his gratitude
to the University of South Africa for its kind hospitality, and the other
(NVS) to the Laboratory of Neutron Physics of JINR (Dubna) for its financial
support.

\begin{table}
\caption{Temperature dependence of the non-radiative capture rate,
$\langle{\cal R}_{e\tau\alpha}\rangle_{\theta}$,
in  units ${\rm cm}^6{\rm mole}^{-2}{\rm sec}^{-1}$.}
\label{trate}
\begin{tabular}{|c|c||c|c|}
\hline
\phantom{-}$\theta\ (10^6\ {}^{\circ}{\rm K})$\phantom{-} &
$\langle \Sigma \rangle N_A^2$ &
\phantom{-}$\theta\ (10^6\ {}^{\circ}{\rm K})$\phantom{-} &
$\langle \Sigma \rangle N_A^2$ \\
\hline
1  & $4.653\times 10^{-55}$ & 100  & $ 2.656\times 10^{-10}$ \\
2  & $3.920\times 10^{-43}$ & 110  & $ 6.177\times 10^{-10}$ \\
3  & $2.363\times 10^{-37}$ & 120  & $ 1.302\times 10^{-9}$ \\
4  & $1.039\times 10^{-33}$ & 130  & $ 2.531\times 10^{-9}$ \\
5  & $4.034\times 10^{-31}$ & 140  & $ 4.606\times 10^{-9}$ \\
6  & $3.797\times 10^{-29}$ & 150  & $ 7.927\times 10^{-9}$ \\
7  & $1.427\times 10^{-27}$ & 160  & $ 1.301\times 10^{-8}$ \\
8  & $2.839\times 10^{-26}$ & 180  & $ 3.119\times 10^{-8}$ \\
9  & $3.549\times 10^{-25}$ & 200  & $ 6.605\times 10^{-8}$ \\
10 & $3.122\times 10^{-24}$ & 250  & $ 2.944\times 10^{-7}$ \\
11 & $2.087\times 10^{-23}$ & 300  & $ 9.123\times 10^{-7}$ \\
12 & $1.121\times 10^{-22}$ & 350  & $ 2.236\times 10^{-6}$ \\
13 & $5.034\times 10^{-22}$ & 400  & $ 4.661\times 10^{-6}$ \\
14 & $1.950\times 10^{-21}$ & 450  & $ 8.636\times 10^{-6}$ \\
15 & $6.671\times 10^{-21}$ & 500  & $ 1.464\times 10^{-5}$ \\
16 & $2.053\times 10^{-20}$ & 600  & $ 3.469\times 10^{-5}$ \\
18 & $1.499\times 10^{-19}$ & 700  & $ 6.853\times 10^{-5}$ \\
20 & $8.289\times 10^{-19}$ & 800  & $ 1.195\times 10^{-4}$ \\
25 & $2.538\times 10^{-17}$ & 900  & $ 1.901\times 10^{-4}$ \\
30 & $3.432\times 10^{-16}$ & 1000 & $ 2.826\times 10^{-4}$ \\
40 & $1.511\times 10^{-14}$ & 1250 & $ 6.170\times 10^{-4}$ \\
50 & $2.208\times 10^{-13}$ & 1500 & $ 1.104\times 10^{-3}$ \\
60 & $1.697\times 10^{-12}$ & 1750 & $ 1.738\times 10^{-3}$ \\
70 & $8.606\times 10^{-12}$ & 2000 & $ 2.507\times 10^{-3}$ \\
80 & $3.273\times 10^{-11}$ & 2500 & $ 4.380\times 10^{-3}$ \\
90 & $1.009\times 10^{-10}$ & 3000 & $ 6.583\times 10^{-3}$ \\
\hline
\end{tabular}
\end{table}
\begin{table}
\caption{Non-radiative reaction rate,
$\langle{\cal R}_{e\tau\alpha}\rangle_{\theta}$, its ratio to the
corresponding radiative reaction rate, and the rate of the energy
generation due to the non-radiative fusion, as functions of the radial
position in the sun.}
\label{rrate}
\begin{tabular}{|c|c|c|c|}
\hline
\phantom{-}$R/R_\odot$\phantom{-} &
\phantom{-}$\langle{\cal R}_{e\tau\alpha}\rangle_{\theta}$
\ cm$^{-3}$sec$^{-1}$\phantom{-} &
\phantom{-}${\langle{\cal R}_{e\tau\alpha}\rangle_{\theta}}^{\mathstrut}/
{\langle{\cal R}_{\tau\alpha}\rangle_{\theta}}_{\mathstrut}$\phantom{-} &
\phantom{-}$\Delta E$ \ erg/(cm$^{3}$sec)\phantom{-}\\
\hline
0.00 & 7148 & $0.294\times10^{-3}$ & $0.182\times10^{-13}$\\
0.01 & 7110 & $0.288\times10^{-3}$ & $0.181\times10^{-13}$\\
0.02 & 6387 & $0.282\times10^{-3}$ & $0.162\times10^{-13}$\\
0.03 & 5416 & $0.272\times10^{-3}$ & $0.138\times10^{-13}$\\
0.04 & 4139 & $0.260\times10^{-3}$ & $0.105\times10^{-13}$\\
0.05 & 3202 & $0.246\times10^{-3}$ & $0.814\times10^{-14}$\\
0.06 & 2194 & $0.230\times10^{-3}$ & $0.558\times10^{-14}$\\
0.07 & 1407 & $0.213\times10^{-3}$ & $0.358\times10^{-14}$\\
0.08 &  810 & $0.196\times10^{-3}$ & $0.206\times10^{-14}$\\
0.09 &  515 & $0.180\times10^{-3}$ & $0.131\times10^{-14}$\\
0.10 &  270 & $0.163\times10^{-3}$ & $0.688\times10^{-15}$\\
0.15 &   10 & $0.962\times10^{-4}$ & $0.258\times10^{-16}$\\
0.20 &0.241 & $0.517\times10^{-4}$ & $0.612\times10^{-18}$\\
0.30 &$0.334\times10^{-4}$ & $0.123\times10^{-4}$ & $0.850\times10^{-22}$\\
0.40 &$0.126\times10^{-9}$ & $0.274\times10^{-5}$ & $0.321\times10^{-27}$\\
0.50 &$0.194\times10^{-14}$& $0.643\times10^{-6}$ & $0.493\times10^{-32}$\\
\hline
\end{tabular}
\end{table}
\begin{table}
\caption{Temperature dependence of the ratio of the three-- to two--body
reaction rates for the primordial nucleosynthesis.}
\label{prim}
\begin{tabular}{|c|c|}
\hline
\phantom{-}$\theta\ (10^6\ {}^{\circ}{\rm K})$\phantom{-} &
\phantom{-}${\langle{\cal R}_{e\tau\alpha}\rangle_{\theta}}^{\mathstrut}/
{\langle{\cal R}_{\tau\alpha}\rangle_{\theta}}_{\mathstrut}$\phantom{-}\\
\hline
300   & $0.181\times10^{-8}$\\
350   & $0.404\times10^{-7}$\\
400   & $0.428\times10^{-6}$\\
450   & $0.275\times10^{-5}$\\
500   & $0.124\times10^{-4}$\\
600   & $0.124\times10^{-3}$\\
700   & $0.671\times10^{-3}$\\
800   & $0.244\times10^{-2}$\\
900   & $0.684\times10^{-2}$\\
1000  & $0.159\times10^{-1}$\\
1250  & $0.766\times10^{-1}$\\
1500  & 0.231\\
1750  & 0.530\\
2000  & 1.02\\
2500  & 2.70\\
3000  & 5.44\\
\hline
\end{tabular}
\end{table}

\begin{figure}
\caption{Jacobi vectors in the configuration and momentum space.}
\label{Jacobi}
\begin{center}
%
\end{center}
\end{figure}
\newpage

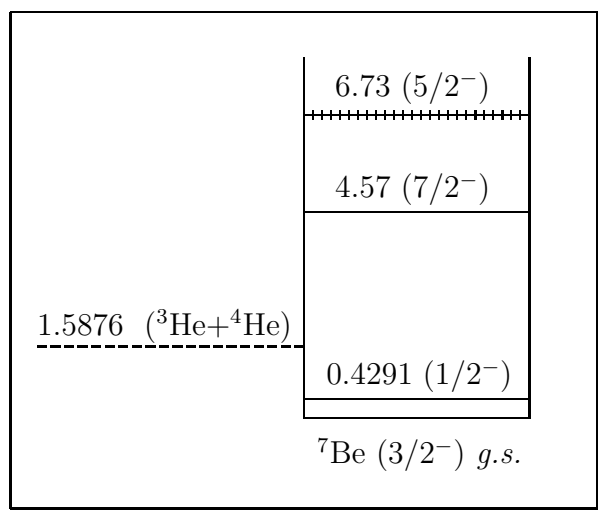
\end{document}